\documentclass[aps,prd,superscriptaddress]{revtex4}
\usepackage{epsfig,epsf}
\usepackage{amsmath}
\usepackage{amsthm}
\usepackage{amsfonts}
\usepackage{amssymb}
\usepackage{dsfont}
\usepackage{multirow}
\usepackage{appendix}
\usepackage{slashed}
\usepackage[active]{srcltx}
\usepackage{psfrag}

\begin{document}

\title{The strong decays of the light scalar mesons $f_0(500)$ and $f_0(980)$
}
\date{\today}
\author{S.~S.~Agaev}
\affiliation{Institute for Physical Problems, Baku State University, Az--1148 Baku,
Azerbaijan}
\author{K.~Azizi}
\affiliation{Department of Physics, Do\v{g}u\c{s} University, Acibadem-Kadik\"{o}y, 34722
Istanbul, Turkey}
\affiliation{School of Physics, Institute for Research in Fundamental Sciences (IPM),
P.~O.~Box 19395-5531, Tehran, Iran}
\author{H.~Sundu}
\affiliation{Department of Physics, Kocaeli University, 41380 Izmit, Turkey}

\begin{abstract}
The partial width of the decay channels $f_0(500) \to \pi \pi$, $f_0(980)
\to \pi \pi$ and $f_0(980) \to K \bar K$ are calculated using QCD light-cone
sum rules method and a technique of the soft meson approximation. The scalar
particles are treated as mixtures of the heavy $|H\rangle =([su][\bar s \bar
u]+[sd][\bar s \bar d])/\sqrt 2$ and light $|L\rangle =[ud][\bar u \bar d]$
scalar diquark-antidiquark components. Obtained results for the full width
of the $f_0(500)$ meson $\Gamma _{\mathrm{th}.}=434.7\pm 72.3~\mathrm{MeV}$
and for the $f_0(980)$ meson $\Gamma _{\mathrm{th.}}=42.12\pm 6.70~\mathrm{MeV%
}$ are compared with the world averages for these parameters, and
a reasonable agreement between them is found.
\end{abstract}

\maketitle

\textbf{1.} Light scalar mesons with masses $m<1\ \mathrm{GeV}$ form a
family of particles, structure and properties of which remain unclear till
now and give rise to different models and theories. The standard model of
the mesons and baryons that considers mesons as bound states of quarks and
antiquarks could not correctly describe the mass hierarchy of these
particles. Therefore, the scalars especially $f_{0}(500)$ and $f_{0}(980)$
mesons have already been in the spotlight of unconventional theories
claiming to solve relevant problems. In most of existing models the light
scalar mesons are treated as multi-quark states: These particles were
considered as four-quark states $q^{2}\bar{q}^{2}$ \cite{Jaffe:1976ig}, or
analyzed as meson-meson molecules \cite{Weinstein:1982gc,Weinstein:1990gu}.
Experimental investigation of the light scalars also meets with
difficulties. Their masses and widths are known with large uncertainties,
which generate additional problems for theoretical studies. Indeed, for
example, the mass and full width of the $f_{0}(500)$ meson is $m=400-550\
\mathrm{MeV}$ and $\Gamma =400-700\ \mathrm{MeV}$ \cite{Patrignani:2016xqp},
respectively. The experimental data of this quality almost do not restrict
suggested models. The contemporary physics of the light scalars embraces
variety of ideas, models and theories, information on which can be found in
the reviews \cite{Amsler:2004ps,Bugg:2004xu,Jaffe:2004ph,Klempt:2007cp}.

The diquark-antidiquark model of the light scalar mesons \cite%
{Jaffe:1976ig,Maiani:2004uc,Hooft:2008we} opened new opportunities for their
theoretical studies. This model was used to calculate the spectroscopic
parameters and width of the scalar mesons in the context of various
computational schemes \cite%
{Ebert:2008id,Latorre:1985uy,Narison:1986vw,Brito:2004tv,Wang:2005cn,Giacosa:2006rg, Chen:2007xr,Sugiyama:2007sg,Kojo:2008hk,Wang:2015uha}. Because within some of
these approaches pure diquark-antidiquark states
did not lead to desired predictions for the parameters of the mesons
different mixing schemes were introduced to evade emerged discrepancies. In
these studies the physical particles were considered as superpositions of
diquark-antidiquarks with different flavor structures \cite{Chen:2007xr}, or
as mixtures of diquark-antidiquarks and conventional $q\bar{q}$ mesons \cite%
{Sugiyama:2007sg,Kojo:2008hk,Wang:2015uha}.

Recently, a suggestion was made to treat the scalar mesons by grouping them
into two nonets with masses below and above $1\ \mathrm{GeV}$ \cite%
{Kim:2017yvd}. In this work the possible mixing of the flavor octet and
singlet states inside of each nonet, as well as mixing between states from
the different nonets was systematically elaborated. In our work \cite%
{Agaev:2017cfz} we treated the mesons $f_{0}(500)$ and $f_{0}(980)$ from the
first nonet of the scalar particles by taking into account the mixing of
flavor octet and singlet diquark-antidiquarks by neglecting, at the same
time, their possible mixing with tetraquarks composed of the spin-1
diquarks. To this end, we used the heavy-light basis
\begin{equation}
|\mathbf{H}\rangle =\frac{1}{\sqrt{2}}\left\{ [su][\overline{s}\overline{u}%
]+[ds][\overline{d}\overline{s}]\right\} ,\ |\mathbf{L}\rangle =[ud][%
\overline{u}\overline{d}],  \label{eq:HLbasis}
\end{equation}%
and introduced the two-angle mixing scheme to get the physical mesons
\begin{equation}
\begin{pmatrix}
|f\rangle \\
|f^{\prime }\rangle%
\end{pmatrix}%
=U(\varphi _{H,}\varphi _{L})%
\begin{pmatrix}
|\mathbf{H}\rangle \\
|\mathbf{L}\rangle%
\end{pmatrix}%
,U(\varphi _{H,}\varphi _{L})=%
\begin{pmatrix}
\cos \varphi _{H} & -\sin \varphi _{L} \\
\sin \varphi _{H} & \cos \varphi _{L}%
\end{pmatrix}%
.  \label{eq:TwoMA}
\end{equation}%
For simplicity in Eq.\ (\ref{eq:TwoMA}), and in what follows we use the
notations $f=f_{0}(500)$ and $f^{\prime }=f_{0}(980)$.

Calculations performed in Ref.\ \cite{Agaev:2017cfz} using QCD two-point sum
rules approach led to the following results for the mixing angles
\begin{equation}
\varphi _{H}=-28^{\circ }.87\pm 0^{\circ }.42,\ \ \varphi _{L}=-\ 27^{\circ
}.66\pm 0^{\circ }.31.\ \   \label{eq:AngRes1}
\end{equation}%
For masses of the scalar particles we obtained
\begin{equation}
m_{f}=(518\pm 74)\ \ \mathrm{MeV,\ \ \ \ }m_{f^{\prime }}=(996\pm 130)\ \ \
\mathrm{MeV,}  \label{eq:MassRes}
\end{equation}%
which are in reasonable agreement with the experimental data.

Apart form the masses of the mesons we defined also their couplings
\begin{equation}
\langle 0|J^{i}|f(p)\rangle =F_{f}^{i}m_{f},\,\ \langle 0|J^{i}|f^{\prime
}(p)\rangle =F_{f^{\prime }}^{i}m_{f^{\prime }},\ \ i=H,L,  \label{eq:Coupl}
\end{equation}%
and suggested that they follow the pattern of state mixing
\begin{equation}
\begin{pmatrix}
F_{f}^{H} & F_{f}^{L} \\
F_{f^{\prime }}^{H} & F_{f^{\prime }}^{L}%
\end{pmatrix}%
=U(\varphi _{H,}\varphi _{L})%
\begin{pmatrix}
F_{H} & 0 \\
0 & F_{L}%
\end{pmatrix}%
.  \label{eq:2AngleCoupl}
\end{equation}%
Here $F_{H}$ and $F_{L}$ can be formally interpreted as couplings of the
\textquotedblleft particles" $|\mathbf{H}\rangle $ and $|\mathbf{L}\rangle $%
. Calculations using QCD two-point sum rules allowed us to evaluate them and
find
\begin{equation}
F_{H}=(1.35\pm 0.34)\cdot 10^{-3}\ \mathrm{GeV}^{4},\ \ \ F_{L}=(0.68\pm
0.17)\cdot 10^{-3}\ \mathrm{GeV}^{4}.  \label{eq:CouplRes}
\end{equation}

In the present Letter we extend our investigation of the $f_{0}(500)$ and $%
f_{0}(980)$ mesons by analyzing a mechanism of their strong decays and
calculate corresponding partial widths. To this end, we use an information
on the $f-f^{\prime }$ system's parameters, i. e. on the masses, mixing
angles and coupling constants, which were extracted from analysis of the
two-point sum rules in Ref.\ \cite{Agaev:2017cfz} and  are not subject
to any adjustments. In investigations we employ QCD light-cone sum rule (LCSR) method \cite%
{Balitsky:1989ry} and technical tools of the soft-meson approximation \cite%
{Belyaev:1994zk}. It is worth noting that these methods were adapted in
Ref.\ \cite{Agaev:2016dev} to study strong vertices composed of tetraquarks
and two conventional mesons.

\textbf{2.} The dominant strong decay channels of the $f_{0}(500)$ and $%
f_{0}(980)$ mesons are the processes $f_{0}(500)\rightarrow \pi \pi $ and $%
f_{0}(980)\rightarrow \pi \pi $. The decay $f_{0}(980)\rightarrow K\overline{%
K}$ was also observed and investigated in experiments \cite%
{Patrignani:2016xqp}. Suggestion on the structure of these scalar particles
as superpositions of the $|\mathbf{H}\rangle $ and $|\mathbf{L}\rangle $
diquark-antidiquark states has important consequences for analysis of their
decays. Indeed, ignoring the mixing phenomenon and assuming that $f_{0}(500)$
and $f_{0}(980)$ mesons are pure $|\mathbf{L}\rangle $ and $|\mathbf{H}%
\rangle $ four-quark states one has to introduce different mechanisms to
describe decays $f_{0}(980)\rightarrow K\overline{K}$ and $%
f_{0}(980)\rightarrow \pi \pi $ : If the first channel runs through the
superallowed Okubo-Zweig-Iizuka (OZI) mechanism, the second one can proceeds
due to one gluon exchange \cite{Brito:2004tv}. The mixing of the $\mathbf{|H}%
\rangle $ and $|\mathbf{L}\rangle $ states to form the physical particles
allows one to treat all of these strong decays on the same footing using the
superallowed OZI mechanism. It is known that the full width of the mesons $%
f_{0}(500)$ and $f_{0}(980)$, which amount to $\Gamma =400-700\ \mathrm{MeV}$
and $\Gamma =10-100\ \mathrm{MeV}$ \cite{Patrignani:2016xqp}, respectively,
suffer from large uncertainties and differ from each other considerably. In
the mixing framework this difference finds its natural explanation: As we
shall see below the dependence of the strong couplings corresponding to the
vertices $f_{0}(500)\pi \pi $ and $f_{0}(980)\pi \pi $ are proportional to $%
1/\sin \varphi _{L}$ and $1/\cos \varphi _{L}$. The dependence of the strong
couplings on the mixing angle $\varphi _{L}$ alongside with other parameters
that enter to sum rules generates a gap in the partial widths of the scalar
particles.

The decay of the $f_{0}(500)$ meson to a pair of pions can proceed through
the processes $f_{0}(500)\rightarrow \pi ^{+}\pi ^{-}$ and $%
f_{0}(500)\rightarrow \pi ^{0}\pi ^{0}$. Let us concentrate on investigation
of the mode $f_{0}(500)\rightarrow \pi ^{+}\pi ^{-}$. In order to calculate
the strong coupling $g_{f\pi \pi }$ we employ QCD light-cone sum rule method
and begin from analysis of the correlation function
\begin{equation}
\Pi (p,q)=i\int d^{4}xe^{ip\cdot x}\langle \pi ^{+}(q)|\mathcal{T}\{J^{\pi
}(x)J^{f^{\dagger }}(0)\}|0\rangle ,  \label{eq:CorrF1}
\end{equation}%
where $J^{f}(x)$ and $J^{\pi }(x)$ are the interpolating currents for the $f$
and $\pi ^{-}$ mesons, respectively. In the two-mixing angle scheme $%
J^{f}(x) $ is given by the formula%
\begin{equation}
J^{f}(x)=J^{H}(x)\cos \varphi _{H}-J^{L}(x)\sin \varphi _{L}.
\end{equation}%
Here $J^{H}(x)$ and $J^{L}(x)$ are the interpolating currents of the scalar
mesons' heavy and light components, respectively. They are defined by means
of the following expressions
\begin{equation}
J^{H}(x)=\frac{\epsilon ^{dab}\epsilon ^{dce}}{\sqrt{2}}\left\{ \left[
u_{a}^{T}(x)C\gamma _{5}s_{b}(x)\right] \left[ \overline{u}_{c}(x)\gamma
_{5}C\overline{s}_{e}^{T}(x)\right] +\left[ d_{a}^{T}(x)C\gamma _{5}s_{b}(x)%
\right] \left[ \overline{d}_{c}(x)\gamma _{5}C\overline{s}_{e}^{T}(x)\right]
\right\} ,  \label{eq:Curr1}
\end{equation}%
and
\begin{equation}
J^{L}(x)=\epsilon ^{dab}\epsilon ^{dce}\left[ u_{a}^{T}(x)C\gamma
_{5}d_{b}(x)\right] \left[ \overline{u}_{c}(x)\gamma _{5}C\overline{d}%
_{e}^{T}(x)\right] .  \label{eq:Curr2}
\end{equation}%
In Eqs.\ (\ref{eq:Curr1}) and (\ref{eq:Curr2}) $a,b,c,d,e$ are color
indices, whereas $C$ is the charge conjugation operator. We interpolate the
pion by means of the pseudoscalar current%
\begin{equation}
J^{\pi }(x)=\overline{u}(x)i\gamma _{5}d(x),  \label{eq:Curr3}
\end{equation}%
with the matrix element defined as%
\begin{equation}
\langle 0|J^{\pi }|\pi ^{-}(p)\rangle =f_{\pi }\mu _{\pi },~\ \ \mu _{\pi }=-%
\frac{2\langle \overline{q}q\rangle }{f_{\pi }^{2}}.  \label{eq:MElem0}
\end{equation}%
In Eq.\ (\ref{eq:MElem0}) $f_{\pi }$ and $\langle \overline{q}q\rangle $ are
the pion decay constant and the quark vacuum condensate, respectively.

The required LCSR can be derived after standard operations: One has to
calculate the correlation function employing physical parameters of the
involved mesons and equate it to an expression of $\Pi (p,q)$ obtained in
terms of the quark-gluon degrees of freedom. We start from the physical
representation of the correlation function $\Pi (p,q)$ that is given by the
formula%
\begin{equation}
\Pi ^{\mathrm{Phys}}(p,q)=\frac{\langle 0|J^{\pi }|\pi ^{-}(p)\rangle }{%
p^{2}-m_{\pi }^{2}}\langle \pi ^{-}(p)\pi ^{+}(q)|f(p^{\prime })\rangle
\frac{\langle f(p^{\prime })|J^{f\dagger }|0\rangle }{p^{\prime 2}-m_{f}^{2}}%
+\ldots ,  \label{eq:Phys1}
\end{equation}%
where $p^{\prime },\ p$ and $q$ are four-momenta of the $f$, $\pi ^{-}$ and $%
\pi ^{+}$ mesons, respectively. The contribution of the exited states and
continuum is denoted in Eq.\ (\ref{eq:Phys1}) by dots. The matrix element of
the pion that enters to this expression is well known. The element $\langle
f(p^{\prime })|J^{f\dagger }|0\rangle $ can be found by taking into account
the structure of the current $J^{f}(x)$ and the fact that only its light
component contributes to this matrix element $\langle f(p^{\prime
})|J^{f\dagger }|0\rangle =F_{L}m_{f}\sin ^{2}\varphi _{L}$. We define the
matrix element corresponding to the strong vertex in the following manner%
\begin{equation}
\langle \pi ^{-}(p)\pi ^{+}(q)|f(p^{\prime })\rangle =g_{f\pi \pi }p\cdot
p^{\prime }.
\end{equation}%
When applying the LCSR method to vertices composed of a tetraquark and two
conventional mesons one has to use a technique of the soft-meson
approximation \cite{Agaev:2016dev}.
The reason is that the tetraquark contains four valence
quarks and contraction with two quark fields from a meson leads to local
matrix elements of the remaining light meson. Then the conservation of the
four-momentum at the vertex requires fulfilment of the equality $q=0$ 
(or $p^{\prime }=p$). In other words, in the case of the tetraquark-meson-meson
vertex the soft-meson approximation is only way to calculate the corresponding
correlation function. For vertices of conventional mesons
the correlation function can be expressed in terms  of a  meson's
distribution amplitudes. This is the full LCSR approach within of which
one may employ the soft approximation, as well. 
For our purposes a decisive fact is the observation made in Ref.\ \cite%
{Belyaev:1994zk}: the soft-meson approximation and full LCSR treatment
of the conventional mesons' vertices leads for  strong couplings to results
that are numerically very close to each other.

In the soft-meson approximation we have to use the  one-variable Borel transformation and
subtract  unsuppressed terms in the physical side of the sum rules. 
We neglect also the mass one of the final
mesons in $\Pi ^{\mathrm{OPE}}(p,q=0)$ and $\Pi _{K}^{\mathrm{OPE}}(p,q=0)$.
Detailed studies of mass effects in exclusive processes prove that they induce only
twist-4 contributions to physical quantities under consideration \cite%
{Agaev:2010aq}. Hence, in the soft-meson approximation the mass effects are also
subleading corrections.

In order to compute the tetraquark-meson-meson vertex we  use the
one-variable Borel transformation, which for the $\Pi ^{\mathrm{Phys}}(p)$
(we use $\Pi (p)\equiv \Pi (p,0)$) leads to the following result%
\begin{equation}
\mathcal{B}\Pi ^{\mathrm{Phys}}(p)=g_{f\pi \pi }f_{\pi }F_{L}\mu _{\pi
}m_{f}m^{2}\sin ^{2}\varphi _{L}\frac{e^{-m^{2}/M^{2}}}{M^{2}}+\ldots ,
\label{eq:Borel1}
\end{equation}%
where $m^{2}=(m_{f}^{2}+m_{\pi }^{2})/2$ and $M^{2}$ is the Borel parameter.
In Eq.\ (\ref{eq:Borel1}) the dots stand for the contribution of the excited
and continuum states, among of which there exist terms that in the soft
limit even after the Borel transformation remain unsuppressed relative to
the ground-state's contribution \cite{Belyaev:1994zk}. In the case under
consideration we are interested only in the ground-state term therefore
these unsuppressed contributions should be removed from Eq.\ (\ref{eq:Borel1}%
). But before performing necessary operations we calculate the $\Pi ^{%
\mathrm{OPE}}(p)$ and find
\begin{equation}
\Pi ^{\mathrm{OPE}}(p)=\sin \varphi _{L}\int d^{4}xe^{ip\cdot x}\epsilon
^{cab}\epsilon ^{cde}\left[ \gamma _{5}\widetilde{S}_{d}^{ib}(x)\gamma _{5}%
\widetilde{S}_{u}^{di}(-x)\gamma _{5}\right] _{\alpha \beta }\langle \pi
^{+}|\overline{u}_{\alpha }^{a}(0)d_{\beta }^{e}(0)|0\rangle .
\label{eq:CurrF2}
\end{equation}%
Computations of $\Pi ^{\mathrm{OPE}}(p)$ using the pion local matrix
elements in accordance with prescriptions explained in rather detailed form
in Ref.\ \cite{Agaev:2016dev}, and the Borel transformation of the obtained
result give
\begin{equation}
\Pi (M^{2})=\frac{f_{\pi }\mu _{\pi }}{16\pi ^{2}}\sin \varphi
_{L}\int_{0}^{\infty }dse^{-s/M^{2}}s+\langle \frac{\alpha _{s}G^{2}}{\pi }%
\rangle \sin \varphi _{L}\frac{f_{\pi }\mu _{\pi }}{16}.  \label{eq:Borel2}
\end{equation}%
In order to perform the continuum subtraction in Eq.\ (\ref{eq:Borel2}) one
has to remove the unsuppressed terms from the $\mathcal{B}\Pi ^{\mathrm{Phys}%
}(p)$ which can be fulfilled by applying the operator \cite{Ioffe:1983ju}
\begin{equation}
\mathcal{P}(M^{2},m^{2})=\left( 1-M^{2}\frac{d}{dM^{2}}\right)
M^{2}e^{m^{2}/M^{2}}.
\end{equation}%
Then for the strong coupling $g_{f\pi \pi }$ we get%
\begin{equation}
g_{f\pi \pi }=\frac{1}{\sin \varphi _{L}}\frac{1}{f_{\pi }F_{L}\mu _{\pi
}m_{f}m^{2}}\mathcal{P}(M^{2},m^{2})\widetilde{\Pi }(M^{2},s_{0}),
\label{eq:StCoupl1}
\end{equation}%
where%
\begin{equation}
\widetilde{\Pi }(M^{2},s_{0})=\frac{f_{\pi }\mu _{\pi }}{16\pi ^{2}}%
\int_{0}^{s_{0}}dse^{-s/M^{2}}s+\langle \frac{\alpha _{s}G^{2}}{\pi }\rangle
\frac{f_{\pi }\mu _{\pi }}{16}.  \label{eq:Borel3}
\end{equation}%
The analysis of the process $f_{0}(500)\rightarrow \pi ^{0}\pi ^{0}$ does
not differ considerably from calculations presented above the difference
being encoded in the current of the $\pi ^{0}$ meson.

\textbf{3. }The decays of the meson $f_{0}(980)$ to $\pi \pi $ and $K%
\overline{K}$ pair proceed by the same superallowed OZI mechanism. In the
case of the process $f_{0}(980)\rightarrow \pi \pi $ the $|\mathbf{L}\rangle
$ component of $f_{0}(980)$ determines the decays $f_{0}(980)\rightarrow \pi
^{+}\pi ^{-}$ and $f_{0}(980)\rightarrow \pi ^{0}\pi ^{0}$. For these
channels a situation does not differ from the decays $f_{0}(500)\rightarrow
\pi \pi $: One needs to replace in Eq.\ (\ref{eq:StCoupl1}) $\sin \varphi
_{L}\rightarrow -\cos \varphi _{L},\ m_{f}\rightarrow m_{f^{\prime }}$, and
set $m^{2}=(m_{f^{\prime }}^{2}+m_{\pi }^{2})/2$. This modifications and
properly chosen parameters $M^{2}$ and $s_{0}$ are enough to perform
numerical analysis of the decay channels $f_{0}(980)\rightarrow \pi ^{+}\pi
^{-}$ and $f_{0}(980)\rightarrow \pi ^{0}\pi ^{0}$, and find their partial
widths.

Investigation of the strong decays $f_{0}(980)\rightarrow K\overline{K}$
actually implies analysis of the following two decay modes $%
f_{0}(980)\rightarrow K^{+}K^{-}$ and $f_{0}(980)\rightarrow K^{0}\overline{K%
}^{0}.$ Naturally, all of these channels run through decays of the $%
f_{0}(980)$ meson's heavy component $|\mathbf{H}\rangle $ . Let us consider
in some details the process $f_{0}(980)\rightarrow K^{+}K^{-}$. The
correlation function necessary to study this decay is%
\begin{equation}
\Pi _{K}(p,q)=i\int d^{4}xe^{ip\cdot x}\langle K^{+}(q)|\mathcal{T}%
\{J^{K}(x)J^{f^{\prime }\dagger }(0)\}|0\rangle ,  \label{eq.CurrF2}
\end{equation}%
where the interpolating current for the $f_{0}(980)$ meson is
\begin{equation}
J^{f^{\prime }}(x)=J^{H}(x)\sin \varphi _{H}+J^{L}(x)\cos \varphi _{L}.
\label{eq:Curr4}
\end{equation}%
For $K$ mesons we use the pseudoscalar current
\begin{equation}
J^{K}(x)=\overline{u}(x)i\gamma _{5}s(x),  \label{eq:Curr5}
\end{equation}%
with the matrix element
\begin{equation}
\langle 0|J^{K}|K^{-}(p)\rangle =\frac{f_{K}m_{K}^{2}}{m_{s}+m_{u}}.
\label{eq.MElem}
\end{equation}%
Skipping details of calculations that are similar to ones presented above we
write down final expressions: Thus, for $\Pi _{K}^{\mathrm{OPE}}(p,q)$ we
get
\begin{equation}
\Pi _{K}^{\mathrm{OPE}}(p,q)=-\sin \varphi _{H}\int d^{4}xe^{ip\cdot x}\frac{%
\epsilon ^{abc}\epsilon ^{dec}}{\sqrt{2}}\left[ \gamma _{5}\widetilde{S}%
_{s}^{ia}(x)\gamma _{5}\widetilde{S}_{u}^{ei}(-x)\gamma _{5}\right] _{\alpha
\beta }\langle K^{+}|\overline{u}_{\alpha }^{b}(0)s_{\beta }^{d}|0\rangle .
\end{equation}%
The final expression for the strong coupling $g_{f^{\prime }KK}$ is%
\begin{equation}
g_{f^{\prime }KK}=-\frac{1}{\sin \varphi _{H}}\frac{m_{s}}{%
f_{K}F_{H}m_{K}^{2}m_{f^{\prime }}m^{\prime 2}}\mathcal{P}(M^{2},m^{\prime
2})\widetilde{\Pi }_{K}(M^{2},s_{0}),  \label{eq:StCoupl2}
\end{equation}%
where $m^{\prime 2}=(m_{f^{\prime }}^{2}+m_{K}^{2})/2$ and
\begin{equation}
\widetilde{\Pi }_{K}(M^{2},s_{0})=\frac{f_{K}m_{K}^{2}}{16\sqrt{2}m_{s}\pi
^{2}}\int_{0}^{s_{0}}dse^{-s/M^{2}}s-\frac{\left( 2\langle \overline{u}%
u\rangle -\langle \overline{s}s\rangle \right) }{12\sqrt{2}}%
f_{K}m_{K}^{2}+\langle \frac{\alpha _{s}G^{2}}{\pi }\rangle \frac{%
f_{K}m_{K}^{2}}{16\sqrt{2}m_{s}}.  \label{eq:Borel4}
\end{equation}%
The strong couplings $g_{f^{\prime }KK}$ and $g_{f^{\prime }K^{0}K^{0}}$
provide necessary information for computing the $f_{0}(980)\rightarrow
K^{+}K^{-}$ and $f_{0}(980)\rightarrow K^{0}\overline{K}^{0}$ decays' widths.

\textbf{4. } In calculations we utilize the light quark propagator (see,
Ref. \ \cite{Agaev:2017cfz}) and use for the quark and gluon condensates the
following values: $\langle \bar{q}q\rangle =-(0.24\pm 0.01)^{3}\ \mathrm{GeV}%
^{3}$, $\langle \bar{s}s\rangle =0.8\ \langle \bar{q}q\rangle $, $\langle
\alpha _{s}G^{2}/\pi \rangle =(0.012\pm 0.004)\,\mathrm{GeV}^{4}$. Apart
from these parameters we also employ the masses of the light quarks $%
m_{u}=m_{d}=0$ and $m_{s}=128\pm 10$ $\mathrm{MeV}$, as well as the masses
and decay constants of the $\pi $ and $K$ mesons: for the pion $m_{\pi ^{\pm
}}=139.57061\pm 0.00024\ \mathrm{MeV}$, $m_{\pi ^{0}}=134.9770\pm 0.0005\
\mathrm{MeV}$ and $f_{\pi }=131\ \mathrm{MeV}$ and for the $K$ meson $%
m_{K^{\pm }}=493.677\pm 0.016\ \mathrm{MeV}$, $m_{K^{0}}=497.611\pm 0.013\
\mathrm{MeV}$ and $f_{K}=155.72\ \mathrm{MeV.}$

For the decays of the $f_{0}(500)$ the working windows for the Borel and
continuum threshold parameters are fixed within the limits%
\begin{equation}
M^{2}=(0.7-1.2)\ \mathrm{GeV}^{2},\ \ s_{0}=(0.9-1.1)\ \mathrm{GeV}^{2}.
\label{eq:Regions}
\end{equation}%
Calculations of the strong couplings lead to the predictions
\begin{equation}
g_{f\pi \pi }=33.94\pm 3.86\ \mathrm{GeV}^{-1},\ \ \ |g_{f\pi ^{0}\pi
^{0}}|=32.76\pm 3.56\ \mathrm{GeV}^{-1}.  \label{eq:Stcoupl3}
\end{equation}%
As a result, for the partial decay width of the processes $%
f_{0}(500)\rightarrow \pi ^{+}\pi ^{-}$ and $f_{0}(500)\rightarrow \pi
^{0}\pi ^{0}$ we find%
\begin{equation}
\Gamma \left[ f_{0}(500)\rightarrow \pi ^{+}\pi ^{-}\right] =223.5\pm 53.7\
\mathrm{MeV,\ \ }\Gamma \left[ f_{0}(500)\rightarrow \pi ^{0}\pi ^{0}\right]
=211.2\pm 48.4\ \mathrm{MeV.}  \label{eq:DW2}
\end{equation}%
The full width of the meson $f_{0}(500)$ is formed almost entirely due to
the decay channel $f_{0}(500)\rightarrow \pi \pi $ because the width of the
mode $f_{0}(500)\rightarrow \gamma \gamma $ is very small. \ It seems
reasonable to compare $\Gamma _{\mathrm{th.}}=434.7\pm 72.3~\mathrm{MeV}$
which is the sum of two partial decay widths (\ref{eq:DW2}) with the
available information on $\Gamma =400-700~\mathrm{MeV}$ noting existence of
an overlapping region of these results. As we have pointed out, data for the
full width of the light scalar mesons suffer from large uncertainties.
Therefore, we can only state that our theoretical prediction is compatible
with experimental data.

\begin{widetext}

\begin{figure}[h!]
\begin{center}
\includegraphics[totalheight=6cm,width=8cm]{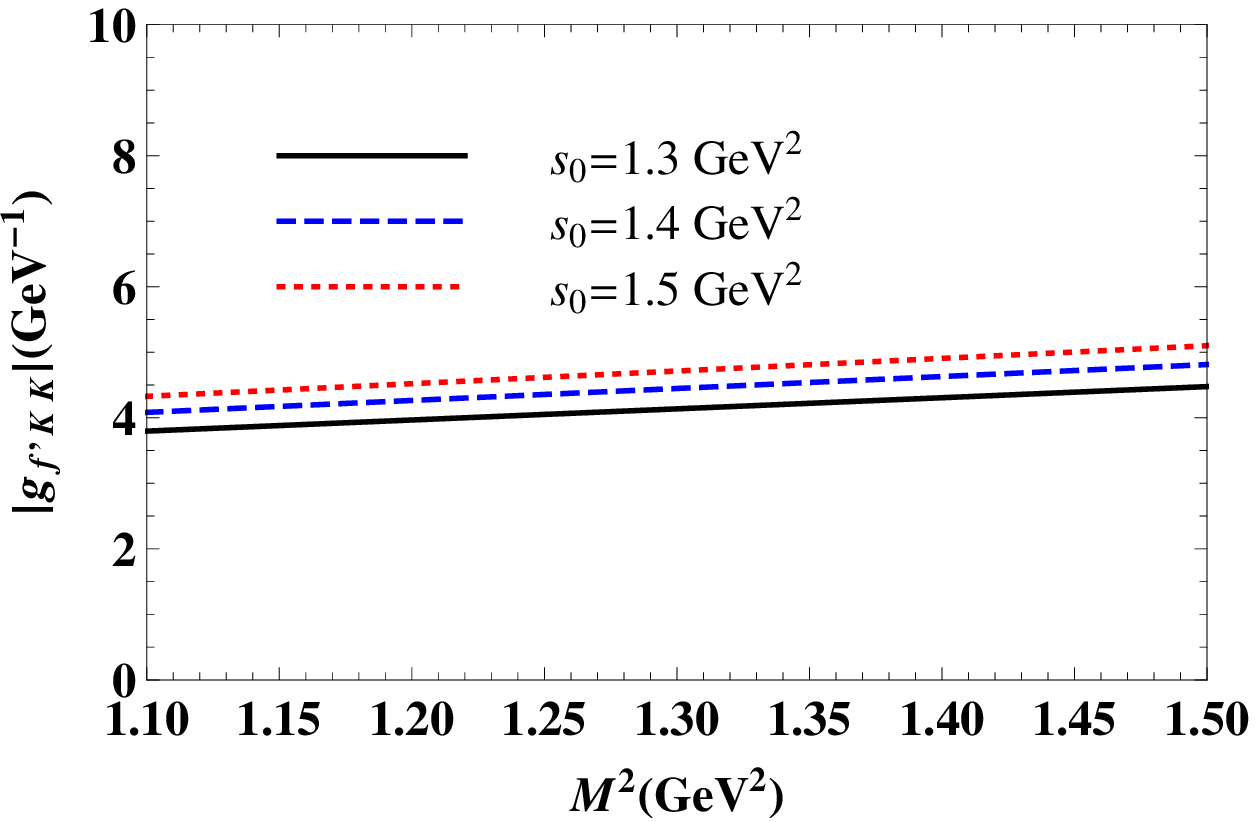}\,\,
\includegraphics[totalheight=6cm,width=8cm]{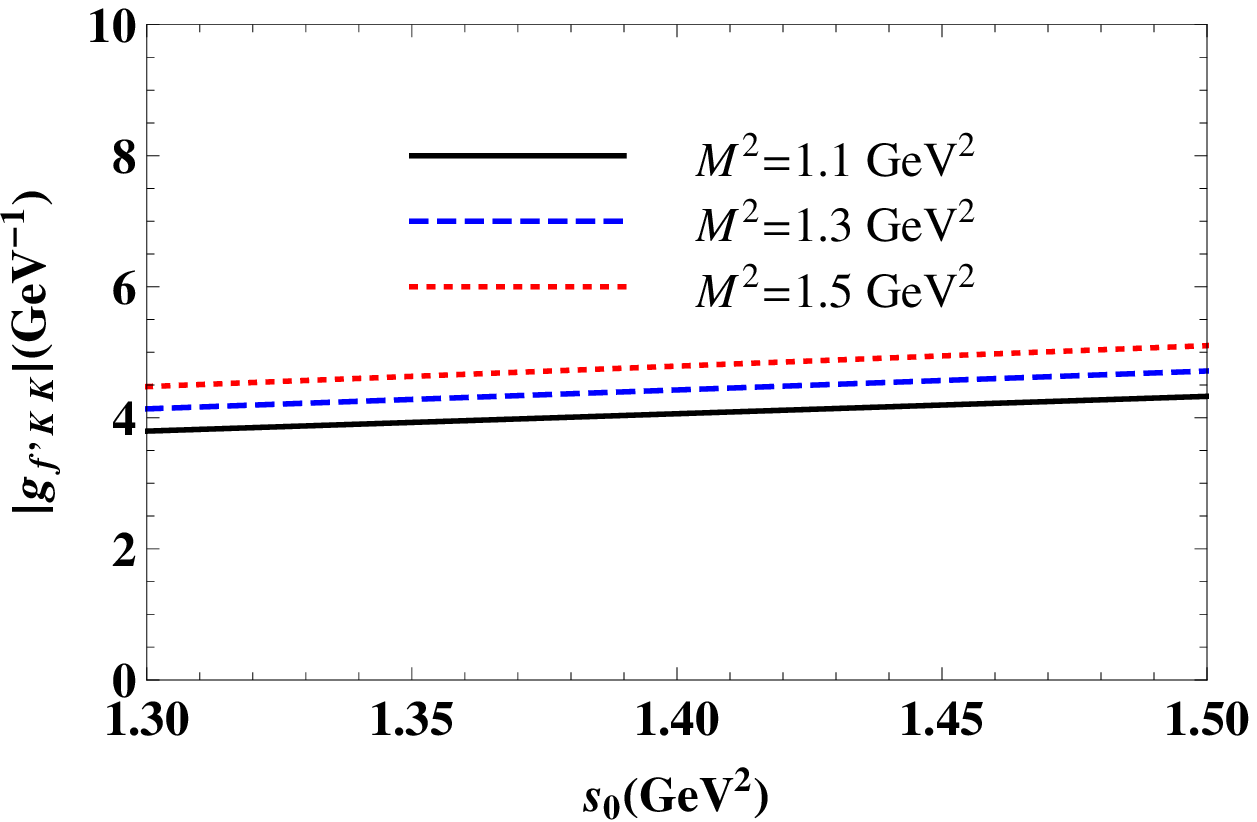}
\end{center}
\caption{ The dependence of the strong coupling $g_{f^{\prime} K K}$  on the Borel parameter $M^2$ at fixed $s_0$
(left panel), and on the continuum threshold $s_0$ at fixed $%
M^2$ (right panel).}
\label{fig:StCoupl}
\end{figure}

\end{widetext}

The strong decays of the $f_{0}(980)$ meson can be analyzed in the same
manner. The differences between the channels  $f_{0}(500)\rightarrow \pi
\pi $ and $f_{0}(980)\rightarrow \pi \pi $ appear due to the spectroscopic
parameters of the involved mesons, and regions chosen for the Borel
parameter and continuum threshold. In the case of the $f_{0}(980)$ meson's
decays we use
\begin{equation}
M^{2}=(1.1-1.5)\ \mathrm{GeV}^{2},\ \ s_{0}=(1.3-1.5)\ \mathrm{GeV}^{2}.
\label{eq:RegionsK}
\end{equation}%
Then for the couplings and partial decay widths we find%
\begin{eqnarray}
g_{f^{\prime }\pi \pi } &=&3.02\pm 0.35\ \mathrm{GeV}^{-1},\ \ \
g_{f^{\prime }\pi ^{0}\pi ^{0}}=3.75\pm 0.45\ \mathrm{GeV}^{-1},  \notag \\
|g_{f^{\prime }KK}| &=&4.29\pm 0.75\ \mathrm{GeV}^{-1},\ \ \ |g_{f^{\prime
}K^{0}K^{0}}|=4.97\pm 0.98\ \mathrm{GeV}^{-1},  \label{eq:StCoupl4}
\end{eqnarray}%
and
\begin{eqnarray}
\Gamma \left[ f_{0}(980)\rightarrow \pi ^{+}\pi ^{-}\right] &=&14.36\pm
3.31\ \mathrm{MeV,\ \ }\Gamma \left[ f_{0}(980)\rightarrow \pi ^{0}\pi ^{0}%
\right] =22.19\pm 5.64\ \mathrm{MeV,}  \notag \\
\Gamma \left[ f_{0}(980)\rightarrow K^{+}K^{-}\right] &=&3.98\pm 1.04\
\mathrm{MeV,\ \ }\Gamma \left[ f_{0}(980)\rightarrow K^{0}\overline{K}^{0}%
\right] =1.59\pm 0.47~\mathrm{MeV}.  \label{eq:DW3}
\end{eqnarray}%
In calculations we have utilized the different working regions for the Borel
parameter $M^{2}$ and continuum threshold $s_{0}$. We have chosen these
regions using standard requirements of the sum rules computations. It is
known that a stability of the obtained results on $M^{2}$ and $s_{0}$ is one
of the important constraints imposed on these auxiliary parameters. We
demonstrate in Fig.\ \ref{fig:StCoupl} as a sample the variation of the
coupling $|g_{f^{\prime }KK}|$ on the $M^{2}$ and $s_{0}$. One can see that $%
|g_{f^{\prime }KK}|$ depends on $M^{2}$ and $s_{0}$, which is a main source
of uncertainties of the evaluated quantities. It is also clear that these ambiguities
are less than $30\% $ of the central values which is acceptable for the sum rule
computations.

It is remarkable that there are valuable experimental information and
independent theoretical predictions for the coupling $g_{f^{\prime }KK}$. It
was extracted from different processes, and calculated by means of numerous
methods. Thus, from analysis of the radiative decay $\phi \rightarrow
f_{0}\gamma $ the CMD-2 and SND collaborations found $g_{f^{\prime
}KK}=4.3\pm 0.5~\mathrm{GeV}$ and $5.6\pm 0.8~\mathrm{GeV}$ \cite%
{Akhmetshin:1999di,Achasov:2000ym}, respectively. The KLOE Collaboration
used the same process and from two different fits extracted the following values $%
g_{f^{\prime }KK}=4.0\pm 0.2~\mathrm{GeV}$ and $5.9\pm 0.1~\mathrm{GeV}$
\cite{Aloisio:2002bt}. Our result for $g_{f^{\prime }KK}$ can be easily
converted to a form suitable for comparison with these experimental data, and
is equal to $4.12\pm 0.72~\mathrm{GeV}$. As is seen, our prediction for
the strong coupling $g_{f^{\prime }KK}$ is in a reasonable agreement with
this experimental information. At the same time, it overshoots experimental
data extracted from other processes such as $D_{s}^{+}\rightarrow \pi
^{-}\pi ^{+}\pi ^{+}$ decay and $pp$ interactions, where the coupling $%
g_{f^{\prime }KK}$ was found equal to $0.5\pm 0.6~\mathrm{GeV}$ and $2.2\pm
0.2~\mathrm{GeV}$ (see, Refs.\ \cite{Aitala:2000xt} and \cite%
{Barberis:1999cq}), respectively.

The theoretical predictions for $g_{f^{\prime }KK}$ appear to vary within wide limits 
and depend on a model accepted for  $f_{0}(980)$ and on methods used in investigations.
For example, in Ref.\ \cite%
{Oller:1997ti} it was found equal to $g_{f^{\prime }KK}=3.8~\mathrm{GeV}$,
whereas in Ref.\ \cite{Colangelo:2003jz} the authors predicted $6.2\leq
g_{f^{\prime }KK}\leq 7.8~\mathrm{GeV}$. The latter estimation was obtained in the
context of the full LCSR method by modeling $f_{0}(980)$ as a scalar meson
with a $\bar{s}s$ component. As it was emphasized by the authors, their
result is larger than previous determinations. It is also larger than our
prediction for $g_{f^{\prime }KK}$ the reason being connected presumably
with a mixing factor of the $\bar{s}s$ component neglected in computations. 
Information on other theoretical studies and references to
corresponding articles can be found in Ref.\ \cite{Colangelo:2003jz}.

Using results presented in Eq.\ (\ref{eq:DW3}) we are able to evaluate the
width of the decays $\Gamma \left[ f_{0}(980)\rightarrow \pi \pi \right]
=36.55\pm 6.54\ \ \mathrm{MeV}$ and $\Gamma \left[ f_{0}(980)\rightarrow K%
\overline{K}\right] =5.57\pm 1.48\ \mathrm{MeV}$. By neglecting the
contribution $\Gamma \left[ f_{0}(980)\rightarrow \gamma \gamma \right] $
for the full width of the meson $f_{0}(980)$ we find $\Gamma _{\mathrm{th.}}=42.12\pm 6.70~\mathrm{MeV%
}$, which is in accord with the experimental data.

\textbf{5. } The partial and full widths of the scalar mesons $f_{0}(500)$
and $f_{0}(980)$ obtained in the present work by treating them as the
mixtures of the different diquark-antidiquark components seem are in
reasonable agreement with existing experimental data. Because there are
great discrepancies between results of different experiments, we
compare our predictions with the world average for these parameters presented by
the Particle Data Group in Ref.\ \cite{Patrignani:2016xqp}. Thus, the full
width of the $f_{0}(500)$ meson is slightly larger than the lower bound of
the experimental data: There is small overlap region between the theoretical
and experimental results. For the $f_{0}(980)$ meson we have found
$\Gamma _{\mathrm{th.}}\in \Gamma _{\mathrm{exp.}}$, which is in a nice agreement 
with the data. Another
parameter $R_{\Gamma }=\Gamma (\pi \pi )/[\Gamma (\pi \pi )+\Gamma
(KK)]=0.87_{-0.08}^{+0.06}$ provides an information on partial decay
widths of the meson $f_{0}(980)$ and on its strange and non-strange
components. The prediction for $R_{\Gamma }$ agrees with the upper
limit for this parameter from Ref.\ \cite{Patrignani:2016xqp}.

As is seen, the model of the light scalar mesons $f_{0}(500)$ and $%
f_{0}(980) $ based on the mixing of the diquark-antidiquark states leads to
the results that are in agreement with the world averages for their full
widths. Nevertheless, some effects which have been neglected in the present
investigation, namely possible mixing with the mesons from the second
(heavier) scalar nonet, as well as $f_{0}(980)-a_{0}(980)$ mixing may
improve our predictions.

The strange and non-strange quark contents of the $f_{0}(500)$ and $%
f_{0}(980)$ mesons also need additional investigations. In fact, the model
accepted here implies that both the mesons $f_{0}(500)$ and $f_{0}(980)$
have the strange and non-strange  components. The existence of sizeable
non-strange content in the $f_{0}(980)$ meson does not contradict to
experimental measurements. But the strange component of the $f_{0}(500)$
meson, as it was pointed out in Ref.\ \cite{Bediaga:2003zh},  may cause
difficulties in interpretation of existing data.
In fact, in Ref.\ \cite{Bediaga:2003zh}  the $f_{0}(500)$
and $f_{0}(980)$ mesons were modeled as mixtures of strange $\overline{s}s
$ and non-strange $(\overline{u}u+\overline{d}d)/\sqrt{2}$ parts. In this 
model the ratio $\Gamma \lbrack
D_{s}^{+}\rightarrow f_{0}(500)\pi ^{+}]/\Gamma \lbrack D_{s}^{+}\rightarrow
f_{0}(980)\pi ^{+}]$ depends on the mixing
angle that has to be extracted from experimental measurements.  
But the E791 Collaboration did not observe a contribution of the process 
$D_{s}^{+}\rightarrow f_{0}(500)\pi ^{+}$  to the decay
$D_{s}^{+}\rightarrow \pi ^{-}\pi ^{+}\pi ^{+}$ \cite{Aitala:2000xt}, which 
contradicts to the theoretical assumption on the strange component of the 
meson $f_{0}(500)$. This experiment predicted  for the strong coupling $g_{f^{\prime }KK}=0.5\pm
0.6~\mathrm{GeV}$, which  contradicts also to all other
measurements. The model used in the present work differs from the framework
introduced in Ref.\ \cite{Bediaga:2003zh}. Therefore, to clarify a situation with 
$f_{0}(500)$ meson's strange component the decays $D_{s}^{+}\rightarrow f_{0}(500)\pi ^{+}$
and $D_{s}^{+}\rightarrow f_{0}(980)\pi ^{+}$ should be studied within this new model.
For comparison to theoretical predictions more precise experimental data are required, as well.

There are no doubts, that the light scalar mesons as unusual particles deserve further detailed
theoretical and experimental studies.

\textbf{6. }K.A. and H.S. thank TUBITAK for the partial financial support
provided under Grant No. 115F183.

\end{document}